\documentclass[]{aa}
\usepackage{graphicx,longtable,lscape}
\usepackage{txfonts,psfig}
\usepackage[]{natbib}
\def \sw {{\it Swift}}

\def\ltsima{$\; \buildrel < \over \sim \;$}
\def\lsim{\lower.5ex\hbox{\ltsima}}
\def\gtsima{$\; \buildrel > \over \sim \;$}
\def\gsim{\lower.5ex\hbox{\gtsima}}

\newcommand{\be}{\begin{equation}}
\newcommand{\en}{\end{equation}}

\begin{document}
  \title{ Swift discovery of the orbital period of the HMXB IGR~J015712--7259 in the Small
  Magellanic Cloud.              }
\author{ A.\ Segreto\inst{1}, G.\ Cusumano\inst{1}, V.\ La Parola\inst{1}, 
A.\ D'A\`i\inst{2}, N.\ Masetti\inst{3}, P.\ D'Avanzo\inst{4}}

   \offprints{G. Cusumano, cusumano@ifc.inaf.it}
   \institute{INAF -- Istituto di Astrofisica Spaziale e Fisica Cosmica di Palermo,
        Via U.\ La Malfa 153, 90146 Palermo, Italy  
\and 
  Dipartimento di Fisica e Chimica, Universit\`a di Palermo, via Archirafi 36, 90123, Palermo,
Italy
\and 
INAF - Istituto di Astrofisica Spaziale e Fisica Cosmica di Bologna, 
via Gobetti 101, 40129, Bologna, Italy
\and
INAF - Brera Astronomical Observatory, via Bianchi 46, 23807, Merate (LC), Italy 
 }

\abstract
{In the last years the hard X-ray astronomy has made a significant step forward, thanks to the
monitoring of the IBIS/ISGRI telescope on board the INTEGRAL satellite and of the Burst Alert 
Telescope (BAT) on board of the Swift observatory. This has provided a huge amount of novel 
information on many classes of sources.}
{We have been exploiting the BAT survey data to study the variability and the spectral 
properties of the new high mass X-ray binary sources detected by INTEGRAL. In this letter 
we investigate the properties of IGR~J015712--7259.}
{We perform timing analysis on the 88-month BAT survey data and on the XRT pointed observations 
of this source. We also report on the broad-band 0.2--150 keV spectral analysis.}
{We find evidence for a modulation of the hard-X-ray emission with period $\rm
P_o=35.6$ 
{\bf d}. The significance of this modulation is  6.1 standard deviations. The broad band spectrum 
is modeled with an absorbed power law with photon index $\Gamma\sim 0.4$ and a steepening in 
the BAT energy range modeled with a cutoff at an energy of $\sim 13$ keV.}
{}
\authorrunning {A.\ Segreto  et al.}
\titlerunning {The orbital period of the HMXB IGR~J015712--7259}

\maketitle

\section{Introduction\label{intro} }

Since the first years of the last decade, astronomy has been having 
two profitable protagonists in the hard X-ray energy band:
the IBIS/ISGRI telescope \citep{ubertini03,lebrun03} on board 
the INTEGRAL satellite \citep{winkler03} and the Burst Alert 
Telescope (BAT, \citealp{bat}) on board of the Swift
observatory \citep{swift}.
IBIS/ISGRI  has conducted a fruitful exploration of the Galactic Plane, 
revealing a large number of new X-ray sources: some of them are
characterized by a strongly absorbed spectrum that made them elusive to 
previous soft X-ray monitoring; 
others show very bright transient episodes and they were revealed thanks to the continuous 
scan of the Galactic Plane.
Many of  these sources  have been identified as high mass X-ray binaries (HMXBs), 
as inferred by the discovery 
of their optical counterparts (e.g., \citealp{filliatre04,reig05, 
masetti06, negueruela06}) and/or by the observation of 
long periodicities due to the occultation of the neutron star by 
the supergiant companion or to the enhancement of the neutron star accretion rate
at periastron passage in an eccentric orbit. 
BAT is playing a momentous role in the study of many of these new INTEGRAL sources
(e.g. \citealp{cusumano10a, laparola10, dai11a, dai11b}).
Thanks to a field of view  two orders of magnitude larger than IBIS/ISGRI and to
frequent changes of pointing direction, it efficiently records
emission variability due to orbital eclipses or to the turn on of transient episodes.

IGR~J015712--7259 is an X-ray binary discovered during the INTEGRAL scan of the
Small Magellanic Cloud (SMC) and of the Magellanic Bridge in December 2008. A Swift-XRT follow-up
observation found the soft X-ray counterpart at RA= 01h 57m 16.4s, 
Dec= $-72^{\circ}$ 58' 33'' (J2000) with an uncertainty localization of 3.8'' 
(90\% confidence level, \citealp{atel1882}), and
the USNO \citep{monet03} star USNO-B1 0170-0064697, that lies within the XRT error box, 
was associated with IGR~J015712--7259. The B and R magnitude of this star are 
15.48 and 15.51, respectively \citep{mcbride10}.
Timing analysis of the XRT and RXTE data revealed a periodicity of 
$\sim 11.6$ s \citep{atel1882}.
A broad band spectral analysis that combines the XRT and ISGRI data
showed a flat power law spectrum ($\Gamma\sim 0.4$) with an exponential cutoff with
a cutoff energy of $\sim 8$ keV \citep{mcbride10}. This spectral shape is consistent
with the spectrum typically shown by a HMXB.

This letter, that reports the results derived by the  analysis of the soft and hard X-ray 
data collected by Swift on IGR~J015712--7259, is organized as follows: section 2 describes the
data reduction;  section 3 reports on the timing analysis;  in section 4 we describe 
the spectral analysis and in section 5 we briefly discuss our results.

\section{Observations and data reduction\label{data}}

\begin{figure}
\begin{center}
\centerline{\includegraphics[width=5.2cm,angle=270]{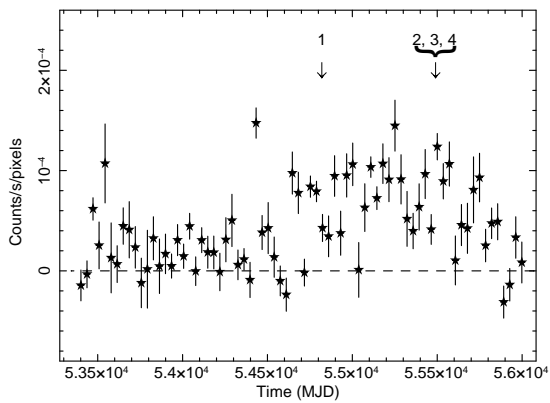}}
\caption[]
{ IGR~J015712--7259 BAT light curve the 15--45 keV energy range, with a  bin 
length of 35 days. 
The epochs of the XRT observations are shown with  vertical arrows.
               }               
                \label{lc} 
\vspace{-1.0truecm}
        \end{center}
        \end{figure}

We have used the  {\sc batimager} code (developed for the analysis of 
coded mask telescopes data, see \citealp{segreto10} for details) to analyze the
data collected by Swift-BAT between November 2004 and March 2012 in survey mode.
IGR~J015712--7259 is detected in the 15--150 BAT all-sky map 
with a significance of $\sim7$ standard deviations in 88 months. 
The light curve of IGR~J015712--7259 (Fig.~\ref{lc}) 
shows that the source is in a low intensity state in the 
first $\sim33$ months, rising to a higher state throughout the following months.
The significance of the source in this second interval rises to $\sim9$ standard
deviation in the 15--150 keV band, and it is maximised to $\sim11$ standard
deviations in the 15--45 keV  energy band.
Therefore, in order to study the timing and spectral properties of the 
source we have used the BAT data collected after MJD 54510.
The light curve with the
maximum resolution achievable with the BAT survey data was extracted in the 15--45
keV energy band. 

The times were corrected to the Solar
System barycentre (SSB) using the task 
{\sc earth2sun}\footnote{http://heasarc.gsfc.nasa.gov/ftools/fhelp/earth2sun.txt}
and the JPL DE-200 ephemeris \citep{standish82}.
The background subtracted spectrum averaged over the entire survey period was 
extracted in eight energy channels and analyzed using the BAT redistribution 
matrix available in the Swift calibration 
database\footnote{http://swift.gsfc.nasa.gov/docs/heasarc/caldb/swift/}.

\begin{figure}
\begin{center}
\centerline{\includegraphics[width=6cm,angle=0]{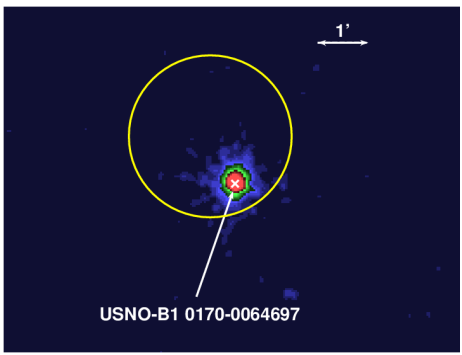}}
\caption[IGR~J015712-7259 sky maps]{ 
0.2-10 keV XRT image with superimposed the position of the optical
counterpart, marked with a cross, and the BAT error circle of 1.61 arcmin 
(yellow circle). 
                }
                \label{map} 
        \end{center}
        \end{figure}

Swift-XRT observed IGR~J015712--7259 four times, once in December 2008 
(ObsID 00031313001) and three
times in October 2010, for a total exposure time of $\sim 12$ ksec. The
details of each XRT observation are in Table~\ref{log}. All the observations are
in Photon Counting mode \citep{hill04}.

\begin{table}
\caption{XRT observation log. \label{log}}
\scriptsize
\begin{tabular}{l l l r r r r }
\hline
    &Obs ID     & $T_{start}$   & $T_{elapsed}$  &Exposure & rate & phase\\
Obs \#    &        &  (MJD)               & (s)            &  (s)    & c/s      &       \\  \hline
1         & 00031313001 & 54820.3171       & 1985.7         & 1950.9  & 0.15     & 0.63  \\
2         & 00041740001 & 55474.6958       & 23255.5        & 4613.9  & 0.02     & 0.03 \\
3         & 00041740002 & 55488.2508       & 4461.2         & 4443.4  & 0.16     & 0.41 \\
4         & 00041740003 & 55490.1957       & 29626.6        & 4214.4  & 0.26     & 0.46 \\
\hline
\end{tabular}
\tablefoot{The quoted phase refers to the profile in
Figure~\ref{period}c. }
\end{table}

Data were  processed with standard procedures, 
filtering and screening criteria  ({\sc xrtpipeline} v.0.12.4).
Fig. 1 shows the 0.2--10 keV XRT image where the soft X-ray counterpart of 
IGR~J015712--7259
is well within the BAT error circle (1.61 arcmin).
We extracted the source events from a circular region (20 pixel radius, with 1 
pixel corresponding to 2.36 arcsec) centered on the source centroid as 
calculated with {\sc xrtcentroid} (RA=01h~57m~15.9s, Dec=$-72^{\circ}$ 58'
29.9'', error radius 3.6'').
The source events arrival times were corrected to the SSB  using the 
task {\sc
barycorr}\footnote{http://http://heasarc.gsfc.nasa.gov/ftools/caldb/help/barycorr.html}.
The background was extracted from an annular
region  with inner and outer radii 40 and 70 pixels, respectively. XRT ancillary 
response files were generated with
{\sc xrtmkarf}\footnote{http://heasarc.gsfc.nasa.gov/ftools/caldb/help/xrtmkarf.html}.
The source and background spectra of each observation were averaged to obtain a
single spectrum, and the ancillary files were combined using {\sc addarf},
weighting them by the exposure times of the relevant spectra. Both the summed 
spectrum and each single spectrum were rebinned with a minimum of 20 counts 
per energy channel, in order to allow the use of the $\chi^2$ statistics.
We used the spectral redistribution matrix v013. The spectral analysis was 
performed using {\sc xspec} v.12.5.
Errors are given at 90\% confidence level, if not stated otherwise.

\begin{figure}[h]
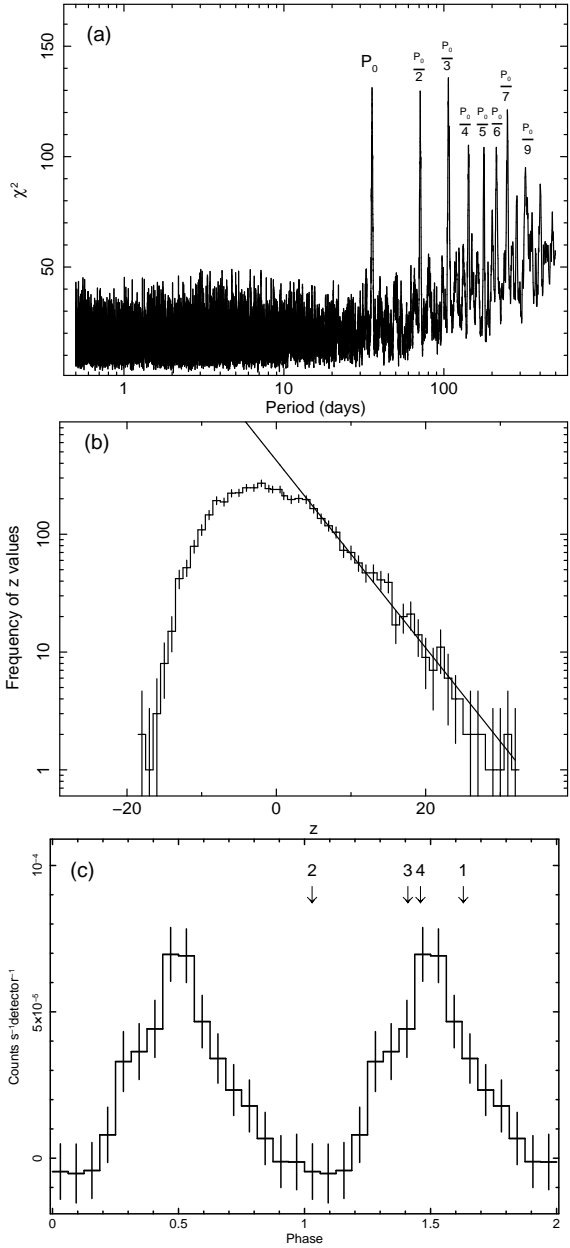

\begin{center}
\centerline{\includegraphics[width=5.5cm,angle=270]{figura2_new.ps}}
\centerline{\includegraphics[width=5.5cm,angle=270]{figura_histo.ps}}
\centerline{\includegraphics[width=5.4cm,angle=270]{figura_profilo.ps}}
\caption[]{{\bf (a)}: Periodogram of \sw-BAT (15--45\,keV) data for 
IGR~J015712--7259.
{\bf (b)}: Histogram of the z ($\chi^2 - F_{\chi^2}$) distribution. The continuous line
represents the best fit exponential model for $z>20$.
{\bf (c)}: Swift-BAT light curve folded at a period {\bf $\rm P_o=35.6$\,d}, with 16 phase 
bins. Two orbital cycles are shown for clarity. The phase relevant to the epoch of each XRT observation is shown with a
vertical arrow.
                }
                \label{period} 
\vspace{-1.5truecm}
       \end{center}
        \end{figure}


        \section{Timing analysis\label{sfxt7:timing}}

The light curve of IGR~J015712--7259 in the 15--45 keV energy range was investigated
for the presence of  periodic modulations.
A timing folding analysis \citep{leahy83} was applied to the baricentered arrival times.
{\bf This method  consists in building  the light curve profile at different
trial periods by folding the photon arrival times in $N$ phase bins. For each
trial profile the $\chi^2$  with respect to the average count rate is
evaluated: a high $\chi^2$ value will signal the presence of a periodic pulsation.}
We searched in the 0.5--500\,d period range with a step resolution 
of $P^{2}/(N \,\Delta T)$, where P is the trial period, $N=16$ is the number 
of folded profile phase bins and $\Delta T$ ($\sim130$ Ms) is the data time span. 
The average rate in each phase bin was calculated weighting the light curve rates 
by the inverse square of the relevant statistical error.
This procedure, adopted to deal with the large span in statistical errors, is
justified by the fact that the data are  characterized by a large range of 
{\bf signal-to-noise ratio}
because BAT monitors the source over a wide range of off-axis directions.
Figure~\ref{period}a shows the resulting periodogram. 
We find several features emerging over the  $\chi^2$ average trend.
The feature at the lowest period is at {\bf $\rm P_{o}=35.6\pm0.5$\,d} 
($\chi^2\sim132$) where the period and its error are evaluated as the
position of the centroid and the standard deviation obtained from a
Gaussian fit of the periodogram feature.
The features at higher P result to be multiples of $\rm P_{o}$.

The average $\chi^2$ in the periodogram increases monotonically with
the trial period deviating from what is expected for a white noise
signal, where the $\chi^2$ is expected to have an average value of $(N-1)$. 
As a consequence, we cannot apply the $\chi^2$ statistics 
to evaluate the significance of the feature at $P_{\rm o}$ and an ad hoc procedure has 
to be used. 
The following steps summarize what we have done to estimate the significance of 
the feature:\\

(1) we modeled the $\chi^2$ distribution with a 2nd order polynomial 
to derive the average trend $F_{\chi^2}(P)$ of the $\chi^2$ versus P. 

(2) $F_{\chi^2}$ was then subtracted from the $\chi^2$ distribution to obtain 
a flattened periodogram (hereafter, z). This allows to evaluate the significance of
the feature with respect to the average noise level of the periodogram. The value of z 
corresponding to $\rm P_{o}$ is $\sim 110$.

(3) We built the histogram of the z distribution (Figure~\ref{period}b) from P=15 d 
to P=55 d (where the periodogram is characterized by a noise level quite consistent with
the noise level $\rm P_{o}$) at excluding the interval around $\rm P_{o}$. 

(4) We modeled the positive tail ($z>5$) of the histogram with an exponential function.
 
(5) We evaluated the area ($\Sigma$) of the z histogram summing the contribution of each single bin 
from its left boundary up to z=5 and integrating the best-fit exponential function beyond
z=5 up to infinity.

(6) We  evaluated the integral of the best-fit exponential function between z=110 and infinity 
and normalized it dividing by $\Sigma$. 

The result ($\sim8.5\times10^{-10}$) is the  probability of chance occurrence to find a z value
$\ge  110$ (or $\chi^2 \ge 132$) and it corresponds to a significance for the $P_o$ feature   
of $\sim6.1$ standard deviations in Gaussian statistics.

In Fig.~\ref{period} (c) we show the BAT light curve folded at  
$\rm P_{o}$ with a $\rm T_{epoch}=55224.8086$ MJD.
The profile is characterized by a single symmetric peak with a minimum consistent 
with null intensity, whose centroid, evaluated 
by fitting the data around the dip with a Gaussian function, is at phase $1.01\pm0.02$
corresponding to MJD (55225.2$\pm0.7$) $\pm \rm nP_{o}$. 
The peak is at phase $1.50\pm0.02$, corresponding to MJD $55242.6\pm 0.7 \pm \rm nP_{o}$. 

The phase corresponding to the epochs of the XRT observations and referred to the folded BAT light curve 
are represented in Figure~\ref{period}c by vertical arrows. The rates averaged over each
observation (column 6 in Table~\ref{log}) show a variability in good agreement with the BAT rate profile.

In order to search for the pulsed modulation at $\sim 11.6$ s we have applied
the folding analysis to XRT pointings 1 and 4 in Table~\ref{log} 
(pointings 2 and 3 have a low statistic
content, with $\sim 70$ counts each). The periodogram derived 
from observation 1 shows a feature at
$\rm P_s=11.58\pm 0.01$ s ($\chi^2 =36.6$, 7 d.o.f.), where 
the error is $\rm P_s^2/(N\Delta T)$, with N=8, confirming the result
reported by \citet{atel1882}. The probability of chance occurrence 
for a feature with such a $\chi^2$
value at 11.6 s is $\sim 8\times 10^{-6}$, corresponding to a significance of $\sim4.5$ standard deviations in
Gaussian statistics. The periodogram of
observation 4 does not show any significant feature. We also performed a folding analysis on each of the
five snapshots of observation 4, without finding any significant feature either in each single 
snapshot periodogram or in the periodogram produced by their sum.

\section{Spectral analysis and results\label{Sanalysis} }

The broad band spectral analysis puts together XRT data of different epochs and BAT data accumulated 
over the 88-month of monitoring. We therefore performed a preliminary analysis to verify that 
no significant spectral variability affects the 4 XRT observations and during the 
BAT monitoring. 
The background subtracted spectra of each single XRT observation were fitted simultaneously
with an absorbed power law with the photon index and absorbing column density forced 
to have the same value for the 
4 datasets. This model, yielding a best fit photon index of $0.99\pm 0.12$ and N$_{\textrm H}$ of
$(1.5\pm0.5)\times 10^{21}$ cm$^{-2}$, 
produced residuals with a similar trend 
for all the datasets. Different BAT spectra were also produced selecting the data in 
different time intervals (MJD intervals 53383.478--54593.061, 54593.061--55233.428, and 
55233.428--56016.101; see Figure~\ref{lc}) and in three different phase intervals 
(0.94--1.31, 0.31--0.44 plus  0.69--0.81, and 0.44--0.69, see Figure~\ref{period}). 
These spectra were fitted with a power law model and, as above, we forced the photon index 
to assume the same value for all the spectra leaving 
the normalization free to vary. The residuals between best fit model and data 
showed  a similar trend for all the datasets. The best fit photon index is $2.0\pm0.2$.

\begin{figure}
\begin{center}
\centerline{\includegraphics[width=6.5cm,angle=0]{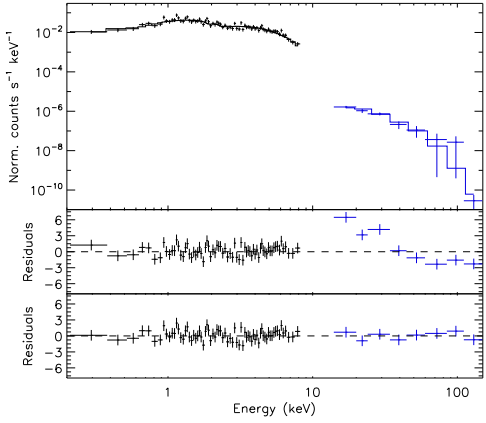}}
\caption[]
{IGR~J015712--7259 broad band spectrum. {\bf Top panel}: 
XRT and BAT data and best fit {\tt phabs*cutoffpl} model.
{\bf Central panel}: Residuals in unit of standard deviations for the {\tt phabs*powerlaw} model.
{\bf Bottom panel}: Residuals in unit of standard deviations for the {\tt phabs*cutoffpl} model.
               }               
                \label{spec} 
\vspace{-1.0truecm}
        \end{center}
        \end{figure}

The broad band spectral analysis was then performed coupling the 15--150 keV 
BAT spectrum extracted from the data collected after MJD 54510 and the XRT 
spectrum obtained by adding the individual
XRT spectra (see Sect.~\ref{data}). 
A multiplicative factor that disengages the normalization parameter of the model for the 
two datasets was introduced in the fit to take into account both the intercalibration 
uncertainty between XRT and BAT and the non simultaneity of the data.
We started fitting the data with an absorbed power law model 
{\tt phabs*(powerlaw)}. 
This model yields an unacceptable $\chi^2$=149.7, with 76 degrees of freedom [dof] 
and is not able to describe the BAT data, as shown in Figure~\ref{spec}, (middle panel). 
The steepening in the BAT energy range 
(see figure~\ref{spec}) suggesting the presence of a cutoff.
Indeed, the spectrum turned out to be much better described adopting the model 
{\tt phabs*cutoffpl} (see table~\ref{fit}) with a photon index  $\Gamma$ of $\sim$ 0.4 
and an folding energy of $\sim$ 13 keV, resulting
in a $\chi^2=70.1$, for 75 dof. The hydrogen column density is lower than  
$\rm 5\times 10^{20} cm^{-2}$.
Figure~\ref{spec} shows data and best fit model (top panel) and residuals 
(bottom panel) for the cutoff powerlaw model.

\begin{table}
\caption{Best fit spectral parameters. \label{fit}}
\scriptsize
\begin{tabular}{ r  l l}
\hline
Parameter        & Cutoff pl                             & Units    \\ \hline 
N$_{\textrm H}$  & $< 5\times 10^{20} $      & $\rm cm^{-2}$\\
$\Gamma$         & $0.4^{+0.1}_{-0.1}$                   &	      \\
$E_{cut}$        &  $13^{+4.0}_{-3}$                & keV\\
$N$              & $5.2^{+0.6}_{-0.4}\times 10^{-4}$     &ph $\rm /(keV~cm^{2}s)$ at 1 keV \\
$C_{BAT}$        & $0.4^{+0.2}_{-0.1}$	         & \\
F (0.2--10 keV)  & $1.26^{+0.11}_{-0.12}\times 10^{-11}$ &$\rm erg~cm^{-2} s^{-1}$ \\
F (15--150 keV)  & $9.1^{+1.5}_{-2.4} \times 10^{-12}$   &$\rm erg~cm^{-2} s^{-1}$ \\
$\chi^2$         &70.1 (75 dof)                          & \\ \hline
\end{tabular}
\tablefoot{$C_{BAT}$  is the constant factor to be multiplied to the
model in the BAT energy range in order to match the BAT data.
We report unabsorbed fluxes for the characteristic XRT (0.2--10 keV) and BAT (10--150 keV)
energy bands. }
\end{table}

\begin{figure}
\begin{center}
\centerline{\includegraphics[width=7cm,angle=0]{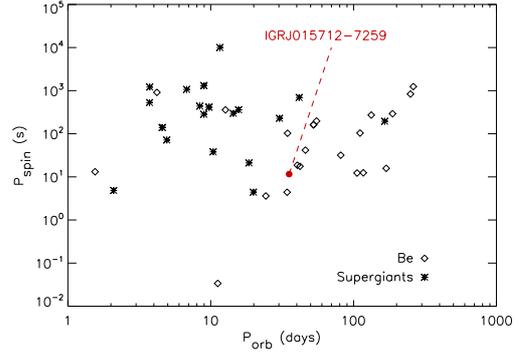}}
\caption[Corbet diagram]{
The Corbet diagram plots the spin period versus the orbital period for HMXBs.
Diamond and star points
represent the Be and supergiant systems, respectively. The red filled circle
marks the position of IGR~J015712--7259.
                }
                \label{corbet}
        \end{center}
        \end{figure}

\section{Conclusions\label{conclusion}}

IGR~J015712--7259, discovered by INTEGRAL in the SMC, is a HXMB with an X-ray emission modulated 
by a spin period of 11.6 s \citep{atel1882}.
The timing analysis of the BAT survey data has allowed to add a new piece of information 
on this binary system, with the discovery of a long term periodic modulation in its hard X-ray 
emission at P$_o=35.6$ {\bf d}. The significance of this result is $\sim 6.1$ standard deviations in 
Gaussian statistics. We interpret this modulation as the orbital period of the binary system.
Its knowledge, together with the spin period measurement allows us to locate the source on the 
Corbet diagram \citep{corbet86}, where its position is consistent with the Be X-ray binaries 
region (Figure~\ref{corbet}).
On the other hand, the BAT light curve folded at P$_o$ shows a triangular symmetric peak with a 
minimum consistent with zero intensity suggesting that accretion happens for most of the orbit. The
minimum could be related to the occultation of the neutron star by the companion star. This
behavior is not typical for HXRB with a Be companion, that usually are observed through 
short-lived enhancements of their emission caused by accretion episodes driven either by 
disk instabilities or to the periastron passage in a highly eccentric orbit. 
                                                    
The BAT long-term light curve (Figure~\ref{lc}) shows that this source  has enhanced its hard X-ray
activity since early 2009, after showing a modest intensity level in the first three years of the
survey monitoring. This behavior prevented the source from being reported in the first Palermo BAT
catalogues \citep{cusumano10b, cusumano10c}, while it is listed with a significance of $\sim 10$
standard deviations in the latest issue of the catalogue\footnote{http://bat.ifc.inaf.it} 
(that covers 66 months of survey). The INTEGRAL detection reported by \citet{atel1882} is located in
time during the early stages of enhanced emission. 

The broad-band (0.2--150 keV) spectral analysis of IGR~J015712--7259 was performed combining all the 
available Swift-XRT observations and the BAT spectrum extracted from the data collected
after MJD 54510 when the source is observed to be in a high intensity state. 
The data are well described by a flat ($\Gamma\simeq 0.4$) powerlaw with a cut-off at 
$\sim 13$ keV. This spectral shape is commonly observed among HMXBs. The best fit parameters are in
agreement with the analysis reported by \citet{mcbride10} based on the December 2008 XRT 
observation and on the ISGRI data.

The results reported above need to be integrated with
optical observations aimed at the identification of the spectral type of the companion star. 
This will allow to ascertain the class of this binary system and to set more constraints on its orbital
characteristics.

\begin{acknowledgements}
This work has been supported by ASI grant I/011/07/0.
\end{acknowledgements}

\bibliographystyle{aa}

{}

\end{document}